# Thermal Control of Size Distribution and Optical Properties in Gallium Nanoparticles

S. Catalán-Gomez[1,2]*, M. Ibáñez[1], J. Rico[1], V. Braza[3], D. F. Reyes[3], M. Villanueva-Blanco[1], E. Squiccimarro[1,#] and J.M. Ulloa[1]

[1]Instituto de Sistemas Optoelectrónicos y Microtecnología, Universidad Politécnica de Madrid, ETSI Telecomunicación, Av. Complutense 30, E-28040 Madrid, Spain

[2]Departamento de Ingeniería Eléctrica, Electrónica Automática y Física Aplicada, Universidad Politécnica de Madrid, ETSI y Diseño Industrial, Ronda de Valencia 3, E-28012 Madrid, Spain

[3]Departamento de Ciencia de los Materiales e IM y QI, Universidad de Cádiz, E-11510 Puerto Real, Cádiz, Spain

*Corresponding author: sergio.catalan.gomez@upm.es

[#]Current address: Department of Applied Science and Technology, Polytechnic of Turin, Corso Duca degli Abruzzi 24, 10129 Turin, Italy

## Abstract

Gallium nanoparticles (Ga-NPs) exhibit promising plasmonic properties spanning ultraviolet to infrared spectral regions, making them suitable for diverse nanophotonic applications. However, the synthesis of uniform and ordered Ga-NP arrays remains challenging due to size heterogeneity arising from coarsening during physical deposition. Here, we systematically investigate the influence of substrate temperature on the nucleation, growth, and homogenization dynamics of Ga-NPs formed via Joule-effect thermal evaporation on GaAs substrates. Atomic force microscopy and scanning electron microscopy reveal temperature-dependent transitions from broad, bimodal size distributions to narrow dispersed arrays within an optimal substrate temperature range of 300-350°C. Above this window, increased diffusion and desorption induce size increase, decreased density, and morphological relaxation manifested as NP shape flattening. Optical reflectance measurements identify distinct localized surface plasmon resonance (LSPR) modes whose energies are closely correlated with NP dimensions and aspect ratios. A figure of merit combining NP density and size uniformity quantifies optimal conditions at intermediate substrate temperatures, consistent with enhanced plasmonic quality factors derived from spectral LSPRs. In-situ post-deposition annealing experiments confirm Ostwald ripening as the dominant coarsening mechanism at elevated temperatures and highlight the necessity of rapid cooling and oxide shell formation to stabilize homogeneous arrays. Cross-sectional transmission electron microscopy with electron energy-loss spectroscopy validates the core-shell structure of individual Ga-NPs, quantifies temperature-dependent oxide shell thickening, and confirms the liquid metal character of the Ga within the core. Finally, size scalability tests with higher deposition times demonstrate that the substrate temperature-driven size homogenization mechanism is valid for a wide range of NP sizes. These results provide a comprehensive framework for thermally tuning the uniformity and optical performance of Ga-NP arrays, advancing their integration into functional plasmonic devices.

## Introduction

Liquid metal nanoparticles (LMNPs) have recently emerged as a hot topic in nanomaterials science due to their unique combination of fluidic behavior, versatile deposition on many substrates by self-assembly, and remarkable physicochemical properties[1-3]. These NPs, which include systems based on gallium (Ga) or their eutectic alloys, exhibit plasmonic properties with low losses across an exceptionally broad spectral range enabling applications in photonics[4], catalysis[5], biomedicine[6] and optoelectronics[7]. One defining feature of LMNPs is their core-shell structure: a liquid metal core encapsulated in a self-limiting oxide shell, which provides environmental stability while providing desirable optical and functional properties[8]. The scalable synthesis of LMNPs has benefited from both top-down and bottom-up methods, allowing control over particle size, shape, and density. This versatility enables LMNPs not only to be deposited on diverse substrates[9] but also to adapt their optical response for use in biosensing, flexible electronics, and other advanced device architectures[10].

Gallium nanoparticles (Ga-NPs), in particular, have attracted attention for their minimal cytotoxicity, broad Localized Surface Plasmon Resonance (LSPR) tunability from ultraviolet to infrared region[11], flexible-dependent properties[12] and versatile core-shell structure[13]. Their unique electronic structure and weak interband transitions contribute to sharp plasmonic resonances and make them compete with noble metals such as Ag and Au[14, 15]. Furthermore, their liquid state also allows morphologically dynamic particles that can be adapted to required shapes and sizes. The ability to precisely control Ga NP dimensions and LSPR is critical for developing optoelectronic and catalytic devices where reproducibility, stability, and performance must be ensured. However, a major challenge persists: deposited-Ga NPs often present broad, heterogeneous size distributions, especially when fabricated on flat surfaces by physical techniques due to growing processes in which two mechanisms are typically ascribed: coalescence (direct physical contact or merging) and Ostwald ripening[16]. This polydispersity leads to inhomogeneous optical responses, making it difficult to reliably implement Ga-NP ensembles in sensing or photonic systems. In contrast, monodisperse and homogeneously ordered NP arrays are desirable[17] for robust, reproducible, and high-performance applications, as heterogeneous ensembles cause broadening of resonances and unpredictable device behavior.

Multiple strategies have been developed to address these limitations and achieve controlled order and homogeneity in Ga-NP arrays. Chemical synthetic routes leverage kinetics and surface ligands to suppress Ostwald ripening and yield monodisperse colloids[18]. Template-assisted methods, such as using nanopatterned aluminum, provide topographic guidance for NP assembly, resulting in highly ordered, close-packed arrays with uniform size and spacing[11]. Physical approaches including ion beam patterning[19] or ion irradiation[20] offer additional control over NP shape and arrangement, enabling the fabrication of non-spherical and size-tunable arrays with improved uniformity. Each method brings distinct advantages and trade-offs in terms of easiness, scalability, substrate compatibility, and achievable order.

Recent research has highlighted substrate temperature as a key parameter for influencing size distribution and morphological evolution during growth on non-templated substrates for physically grown Ga-NPs [16]. It has been demonstrated by in-situ experiments that increasing deposition temperature enhances diffusion, driving coarsening which can reduce polydispersity and shape heterogeneity. While this approach did not achieve perfect monodispersity, it offers a practical route to improving the uniformity of Ga NP ensembles manufactured via physical processes.

While prior studies have demonstrated temperature-enhanced surface diffusion and coarsening in Ga nanodroplets on $SiN_x$ substrates using in-situ scanning transmission electron microscopy (STEM), achieving only partial polydispersity reduction under low-density conditions, our work addresses critical gaps for scalable plasmonic applications on technologically relevant GaAs substrate. We provide the first comprehensive mapping of substrate temperature-driven morphological transitions from heterogeneous bimodal ensembles to dense, more uniform arrays at 300-350 °C, followed by flattening at higher substrate temperature. All the morphological results have been quantified via a novel figure of merit (FoM = density/size FWHM) that optimizes uniformity for collective plasmonics. Unlike template-assisted or chemical routes requiring complex lithography or ligands, this Joule-evaporation approach on flat substrates yields reproducible LSPR quality factors approaching single-particle values (Q ≈ 0.84), validated by STEM core-shell analysis and scalability to larger sizes. These advances enable practical, ligand-free fabrication of homogeneous Ga-NP arrays with tunable optical performance providing practical insights for the scalable fabrication of functional plasmonic nanomaterials.

## Materials and Methods

Ga-NPs were grown by thermal evaporation onto GaAs (001) n+ substrates. High-purity gallium (99.9999%) was evaporated in a vacuum chamber at a base pressure of $3 \cdot 10^{-6}$ mbar using a tungsten crucible operated at 50-60 W. The substrate temperature was changed from RT to 400 °C by applying DC bias to the ceramic sample holder up to 28 V. The actual temperature was measured by a thermocouple placed in the sample holder within the chamber. The NPs radius was controlled by adjusting evaporation time using a shutter placed between crucible and target substrate. 80 s of evaporation time was selected for the primary study to yield NPs where both transversal and longitudinal LSPR modes are detectable (200-900 nm), facilitating direct correlation of spectral splitting with temperature-dependent morphology. Size scalability was tested by changing deposition time from 10 to 150s. After completing the deposition, the substrate heater was switched off and the chamber was vented to air within a few minutes; no deliberate cooling step in vacuum was applied, so samples grown at elevated substrate temperature were exposed to air while still above RT forming a thin native oxide shell around the liquid Ga core.

Surface morphology of Ga-NPs deposited for 80s was studied by Atomic Force Microscopy (AFM) in tapping mode to obtain high-resolution topographic images. The AFM setup was Dimension Icon (Bruker, US). The commercial tapping mode Si cantilevers with a nominal tip radius of ≈8 nm, model ACTA-50, were purchased from AppNano, with a nominal resonance frequency in the range of 200-400 kHz and a nominal force constant in the range of 13-77 N/m. The images were analyzed with Gwyddion software to generate histograms of NP radii and to extract Ga total volume deposited.

Surface morphology of Ga-NPs was also analyzed using a FEI Inspect F50 scanning electron microscope (SEM) operated at 5 keV. Top-view images were acquired to assess the surface of the NPs. Size distribution histograms were extracted from SEM images using the ImageJ software.

Cross-sectional samples for STEM analysis were prepared through conventional polishing, followed by a final thinning step using a Gatan Precision Ion Polishing System (PIPs). Structural characterization and nanoscale compositional analysis were conducted on an FEI Talos F200X STEM operating at 200 kV. To examine NP morphology, high-magnification imaging was performed in annular dark-field (ADF) modes. Elemental composition was analyzed via electron energy loss spectroscopy (EELS), using a Gatan Continuum spectrometer. Data processing was carried out using Velox® and DigitalMicrograph® software.

Optical reflectance spectra were recorded using a Jasco V-670 UV-Vis-NIR spectrophotometer at an incident angle of 8° in 2x2cm samples. Measurements were performed in the 200-900 nm wavelength range.

## Results and Discussion

Figure 1 presents a comprehensive analysis of the morphological and statistical evolution of hemispherical Ga-NPs deposited for 80s by Joule-effect thermal evaporation on a flat GaAs surface varying substrate temperatures from RT to 400 °C. The figure includes atomic force microscopy (AFM) images and corresponding histograms after Gwyddion analysis for NP height and diameter distribution with their corresponding Gaussian fitting, enabling quantitative insight into the impact of substrate temperature on NP assembly and size control.

At RT and 100°C, Ga-NPs AFM images (Figure 1 a) and b)) appear as a dense, heterogeneous assembly with a broad, bimodal size distribution. The histograms show distinct populations of both small and larger NPs with NP mean diameters of 15-16 and 110-115 nm, respectively, highlighting the polydispersity characteristic. This indicates uncontrolled coarsening and limited adatom mobility on the substrate, resulting in numerous small particles alongside larger ones. For the sake of comparison for the following temperatures, it is worth noting that population of small NPs presents higher density than the second size mode and that the NP mean heights of this NP population are around 54 nm resulting in almost perfectly hemispherical shapes. Those results are in agreement with a physical deposition without enhanced surface diffusion as reported previously for RT growth in different flat surfaces such as Si, sapphire, glass and also GaAs[7, 9]. Upon increasing the substrate temperature to 200°C (Fig. 1 c)), the NP populations shift toward larger mean heights but still maintain a bimodal distribution. This reflects the initiation of surface diffusion-driven coarsening, with small NPs beginning to shrink or disappear. In order to follow the NP diameter and height evolution of the big NPs (second mode of the bimodal size distribution), both parameters have been plotted as a function of substrate temperature in Figure 2 a) and b). Error bars for each temperature point are also included and correspond to the Gaussian fit uncertainty. It is worth noting that lateral dimensions from AFM may exhibit minor tip convolution (~8 nm nominal radius), primarily affecting and overestimating densely packed small NPs (first Gaussian distribution mode, excluded from aspect ratio analysis). Nevertheless, heights are unaffected. Dominant population means (≥80 nm) exceed tip scale, with trends corroborated by SEM and TEM in the following figures.

The most prominent transition occurs from 200 to 300 °C and 350 °C. At those temperatures, AFM images (Fig. 1 d) and e)) reveal a profound morphological reorganization: small NPs have vanished, and the NP ensemble converges toward a narrow, quasi-unimodal distribution. The corresponding histograms at these temperatures display mainly a pronounce (high NP density) single peak, demonstrating the emergence of homogeneity in NP size and an overall reduction in polydispersity. Although the histograms at 300 and 350 °C also show a shoulder at 16-30 nm, it is likely negligible since the NP density (vertical axis of nº of NPs/$\mu m^2$) is lower than 2 compared to 14 NPs/$\mu m^2$ of the main NP population (80% reduction). This transition is a direct consequence of thermally activated surface diffusion, which accelerates mass transfer from small to larger NPs, thereby promoting uniform growth and reducing heterogeneity. Interestingly, there is not only a meaningful change in NP distribution but also in absolute diameter and height. Specifically, the NP main diameter at 300 and 350 °C decreases respect to the big NPs at lower substrate temperatures from 100-120 nm to 80 nm as observed in Fig. 2 a) and b). The cause behind this decrease is the total Ga volume deposited. As demonstrated previously[16], for substrate's temperature above 300 °C, Ga sticking is not as efficient as for lower temperatures and a desorption process is apparently produced. Consequently, less Ga total mass is deposited at moderate substrate temperatures such as 300 and 350 °C. After Gwyddion analysis of the AFM images of Fig. 1, the total volume of Ga in NPs at all temperatures in this study has been extracted and plotted as a function of temperature in Fig. 2 c) showing a big step at the expected temperature. Although more studies are needed to understand the process of desorption, we believe that higher desorption provokes lower average NP height and diameter, at the same time than the size distribution is narrowed produced by the increased coarsening due to substrate temperature driven diffusion.

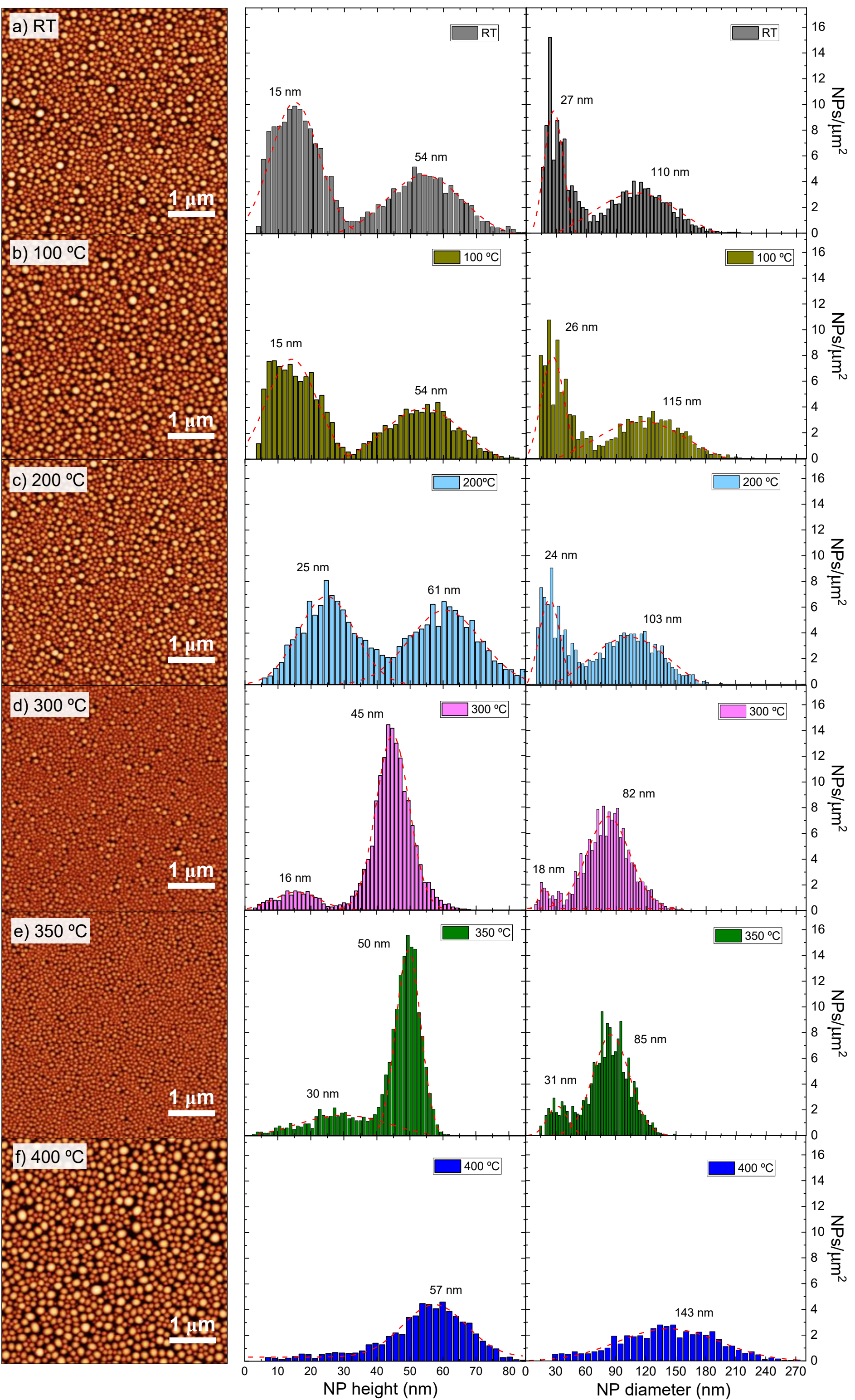

*Figure 1. Morphological characterization of Ga-NPs at varying substrate temperatures. a) to f) AFM images of Ga-NP arrays deposited on GaAs substrates by Joule-effect thermal evaporation at temperatures ranging from RT to 400°C. Accompanying histograms show the distributions of NP height and diameter obtained from Gwyddion analysis, with Gaussian fits illustrating the size populations.*

In Fig. 1 f), at the highest temperature (400°C), there are also meaningful changes: the NP height and diameter increase again, but the distributions broaden slightly although a unique gaussian fit is needed. As a result of the size increase but maintaining almost the same Ga volume fraction (Fig. 2 c)), the density decreases abruptly to have about 14-16 NPs/$\mu m^2$ at 300-350 °C to have only 4-5 NPs/$\mu m^2$ at 400 °C in the same area as observed in height histograms from Fig. 1 d-f).

The evolution of NP shape has also been studied as a function of substrate temperature by analyzing the aspect ratio (height/diameter) of the larger Ga-NPs, as depicted in Fig. 2d). These values have been calculated from Gaussian fit of the histograms. Specifically, from the average height in second column of Fig. 1 divided by average diameter in third column. Error bars are included from Gaussian error propagation in each fitting. Thus, these aspect ratio values are representative of the big NPs population. Observing Fig. 2 d), there are noticeable changes as a function of substrate temperature. At low temperatures (RT to 100°C), the aspect ratio remains relatively constant in the range of 0.47-0.5, indicating that the NPs maintain nearly hemispherical geometries. However, an abrupt change is observed at 200-350°C, where the aspect ratio reaches its maximum value of approximately 0.59, reflecting the formation of more cone-like NP shapes. Notably, at the highest temperature investigated (400°C), the aspect ratio decreases sharply to about 0.40, meaning a significant flattening of the NPs. For the sake of visualization, in Figure S1 are included height line profiles of three NPs grown at RT, 350 °C and 400 °C with their corresponding AFM image indicating the chosen NP. Although the three NPs being the same height, 400 °C NP have a higher diameter, followed by the RT one and finally the 350 °C NP that have the lowest diameter and thus, highest aspect ratio.

This second abrupt change of aspect ratio at 400 °C is consistent with a morphological relaxation process in which Ga-NPs minimize their surface and interfacial energy, likely facilitated by the increased atomic mobility. Such flattening is in line with the well-known tendency of LMNPs to adopt shapes governed by energetic minimization when thermal energy is sufficient to overcome shape anisotropy[3, 21]. This shape transition confirms the strong influence of substrate temperature not only on NP size but also on their three-dimensional morphology, an important factor for their plasmonic and functional properties.

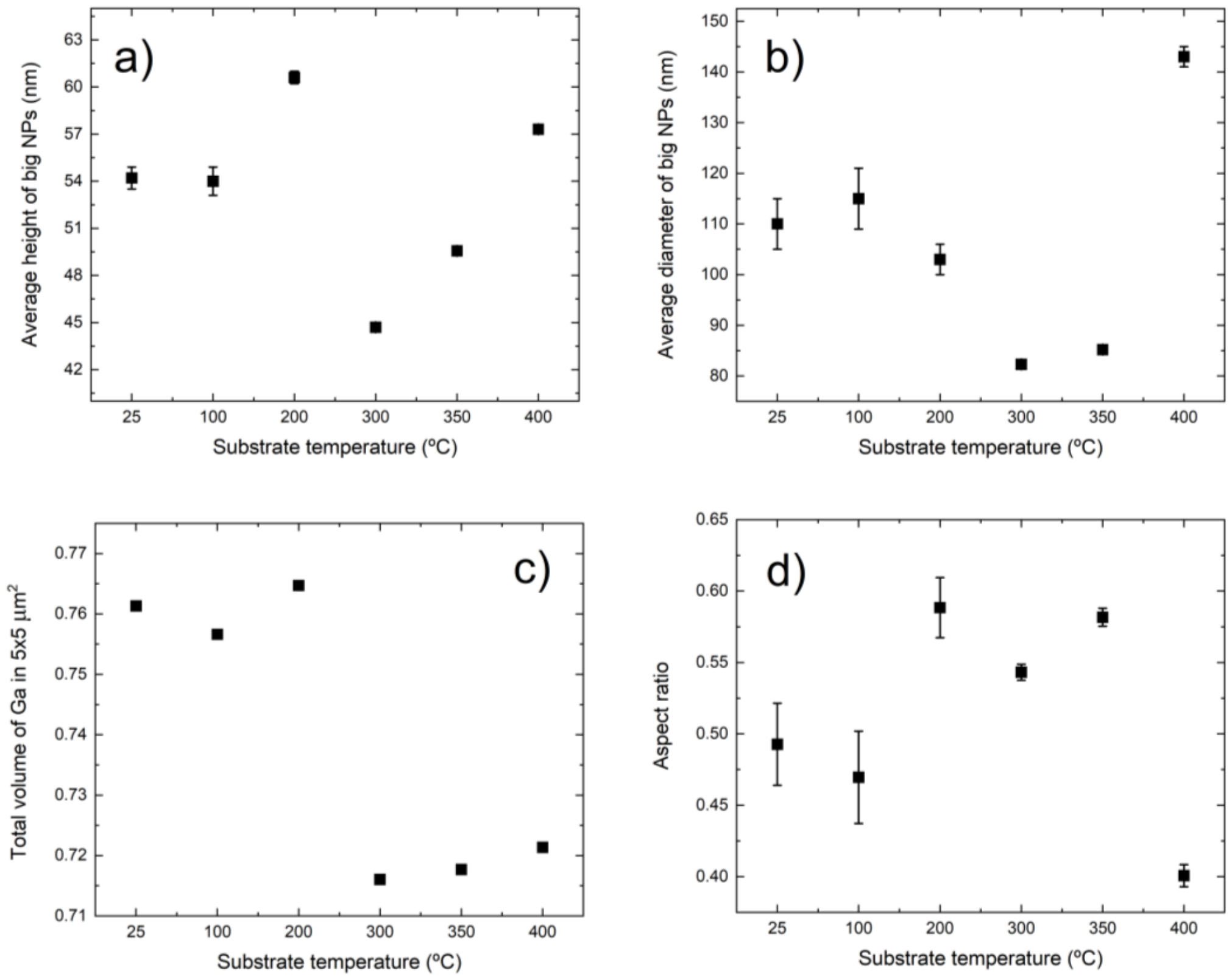

*Figure 2. Quantitative analysis of Ga-NP size metrics and shape evolution with substrate temperature from AFM images of Figure 1. (a) Average NP height, (b) average diameter, (c) calculated total surface volume of Ga and (d) aspect ratio (height divided by diameter) of the larger NP population derived from Gaussian fits of the histograms in Figure 1. Error bars indicate fitting uncertainties.*

In the field of LMNP growth, there is ongoing debate regarding the dominant mechanism: whether NP coalescence occurs by direct physical contact and merging when droplets touch, or whether the evolution is instead governed by Ostwald ripening[22, 23]. The latter is a process in which smaller NPs dissolve and their material diffuses across the substrate to larger ones, which grow at the expense of the disappearing smaller particles[24]. In the Ostwald ripening scenario, mass transport is mediated by surface diffusion rather than direct particle-particle contact, and this process is particularly pronounced at elevated temperatures, where atomic mobility is high[25]. As smaller NPs possess higher surface curvature and thus higher chemical potential, their atoms are less stable and tend to detach, migrate over the substrate or environment, and be incorporated into larger, more stable NPs. This results in a progressive reduction in the population of small NPs and an increase in the size of the larger NPs, leading over time to a more monodisperse array. In-situ TEM measurements during Ga evaporation[16] have confirmed that the Ostwald ripening process is more likely than direct coalescence at elevated temperatures. However, experimental conditions such as substrate type and NP density were different in that work. Thus, to clarify which mechanism predominates under our conditions, we conducted an experiment whose results are shown in Figure S2. Ga-NPs were deposited at substrate temperature of 350°C for 80s (same conditions than in Fig. 1), after which the sample was held at the same temperature for 15, 30 and 50 min within the vacuum chamber with the Ga source closed, meaning no additional material was supplied. This situation allows us to observe the post-growth evolution of the NP ensemble in the absence of extra Ga flux. The size of the NPs for 80s at 350 °C is relatively large (mean diameter ≈85 nm), making it highly unlikely for them to migrate laterally and collide to undergo direct coalescence especially considering the lack of new nucleation events. The results in AFM images of Fig. S2 reveal a clear change in the NP array: after this in-situ annealing, the smaller NPs decrease in number and effectively disappear, while the larger NPs increase in size up to 147 nm, as evidenced by the corresponding height and diameter histograms (Fig. S2 b), c) and d)). This supports the conclusion that Ostwald ripening dominates the growth process under these conditions, since material from the dissolving small NPs is redistributed via surface diffusion to feed the growth of the larger ones. The absence of new nucleation and the likely physical immobility of the NPs rule out direct coalescence as a plausible explanation. Therefore, to reliably achieve the homogeneous scenario observed at 300 °C and 350 °C, it is essential to promptly decrease the substrate temperature at the end of the Ga deposition. By extracting the sample and exposing it to ambient air, the formation of a self-limiting gallium oxide shell is triggered, which surely stabilizes the NP array and effectively inhibits further surface diffusion. This protective oxide layer preserves the optimized NP size distribution and morphology, locking in their uniformity.

Cross-sectional STEM combined with EELS was performed to validate AFM findings and provide direct evidence for the core-shell structure of Ga-NPs for Ga deposition time of 30s at contrasting substrate temperatures (RT versus 350 °C). Figure 3 presents ADF STEM images alongside EELS elemental mapping. At RT deposition conditions (column a), the low-magnification ADF image (a1) reveals broader NP size distribution consistent with the bimodal size distribution from AFM in Fig. 1 a). Higher magnification (a2) clearly visualizes the core-shell structure of the NPs. EELS analysis (a3) confirms a Ga-rich core (green) encapsulated by an oxide shell (orange) at the periphery, while As (blue) is localized only in the substrate, demonstrating interface integrity. The oxide shell thickness has been measured to be in the range of 2.0-3.0 ± 0.5 nm.

NPs deposited at substrate temperature of 350 °C (Fig. 3 b)) reveal notably different morphologies. As shown in the low-magnification ADF image (b1), there has been a substantial enhancement in size homogeneity, thereby directly corroborating the AFM-derived unimodal distribution depicted in Fig. 1 e). Higher magnification images (b2) show that these NPs manifest quasi-spherical geometries in comparison to RT ones, exhibiting higher aspect ratios that are in agreement with the AFM data presented in Fig. 2 d). EELS mapping (b3) also reveals the core-shell configuration albeit with a thickened oxide shell of 4 ± 0.5 nm. The observed shell thickening at 350 °C is likely attributable to enhanced oxidation kinetics during the sample extraction from the evaporation chamber. Seemingly, when freshly deposited, NPs retain significant thermal energy, thereby accelerating the incorporation of oxygen into the oxide layer. Contact angle measurements derived from high-resolution ADF images yielded values of ≈80° at RT and ≈96° at 350 °C, confirming a more pronounced wetting behavior at lower temperatures. These values are consistent with the AFM aspect ratio trends (Fig. 2 d)), where the highest aspect ratios (0.59) are observed at temperatures ranging from 300 to 350 °C.

Remarkably, the core-shell structure is maintained under both substrate temperature conditions, confirming that Ga-NPs retain their liquid character encased within a self-limiting oxide shell. The oxygen-to-gallium stoichiometry in both samples is found to be almost identical (approximately 70% Ga and 30% O), indicating that oxide composition is independent of deposition temperature (see Figure S3.). The STEM-EELS results provide direct nanoscale evidence that supports the AFM statistics, proving that temperature modulates the NP morphology and size distribution through surface diffusion kinetics while preserving the core-shell structure that is critical for Ga-NP functionality.

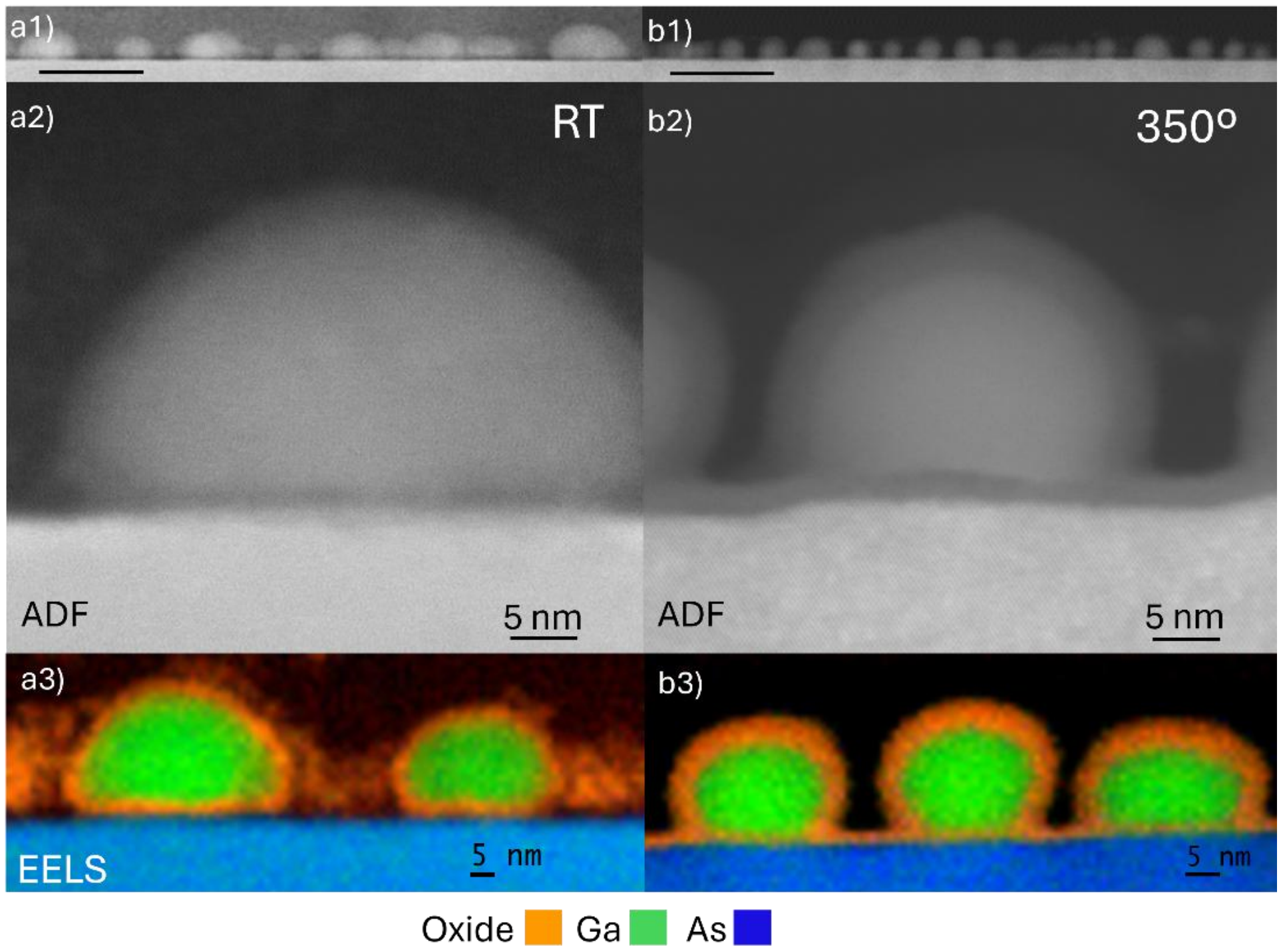

*Figure 3. STEM analysis of both samples where column a) corresponds to RT sample while column b) to the 350° substrate deposited sample. Rows (1) and (2) show low-magnification (scale 100 nm) and high-magnifications ADF images, respectively. In row (3) EELS elemental maps showing the chemical distribution of Ga (green, core), Oxide (orange, shell), and As (blue, substrate).*

Driven by the complex morphological and statistical changes observed with varying substrate temperatures in Fig. 1 and 2 such as shifts in NP density, transitions between bimodal and quasi-unimodal size distributions, narrowing or broadening of population widths, and variations in aspect ratio, it became essential to distill these outcomes into a single, application-relevant metric. For plasmonic and optical purposes, an ideal scenario would comprise a large number of NPs that are highly uniform in size, ensuring a strong, collective, and reproducible optical signal. Therefore, a figure of merit (FoM) was constructed to quantitatively classify the degree of NP homogeneity achieved at different substrate temperatures as followed:

$$FoM = \frac{NPs\ density}{size\ FWHM} \qquad \text{Equation 1}$$

Where FoM is defined as the NPs density (NPs/µm$^2$) divided by the width (FWHM) of the size distribution extracted from Gaussian fitting of the diameter histograms in Fig. 1. The FoM thus reflects not only how many NPs are present, but also how similar they are in size. A high FoM indicates a dense array of nearly identical NPs, while a low FoM means lower density and/or greater polydispersity. Figure 4 shows the calculation of this FoM for the data of this study. In case of two size distribution modes (low temperatures), the FoM have been calculated for the second mode in the histograms since biggest NPs typically determine the optical main signal due to their larger NP cross-section. We have classified three different scenarios as a function of substrate temperature.

At low substrate temperatures (25-200°C) the FoM is low, indicating a sparse distribution and high heterogeneity. The histograms from Fig. 1 show broad, bimodal distributions, NPs vary widely in size, and the array is non-homogeneous. The schematic inset in Fig. 4 illustrates this as randomly arranged, differently sized NPs. This regime is typical for physical deposition at low substrate temperatures, where limited adatom mobility results in poor control over nucleation and growth. As substrate temperature increases (300-350 °C), the FoM rises dramatically and reaches its maximum at approx. 350 °C since the density of NPs rises while the FWHM of the size distribution decreases. This means there are many NPs and crucially, they are highly uniform in size. The included sketch shows as closely packed, similarly sized NPs. This is due to enhanced surface diffusion and Ostwald ripening, which favors growth of uniform NPs at the expense of small, unstable ones. This scenario is ideal for reproducible collective optical properties. Lastly, at the highest temperature (400 °C), the FoM decreases again. Since there is an increase in average NP size, the density drops (fewer NPs/μm²), and the size distribution broadens (FWHM increases). The schematic shows larger, less uniform blue NPs, marking the loss of optimal uniformity. At this stage, the mobility and desorption of Ga are so pronounced that maintaining both high density and uniformity becomes difficult.

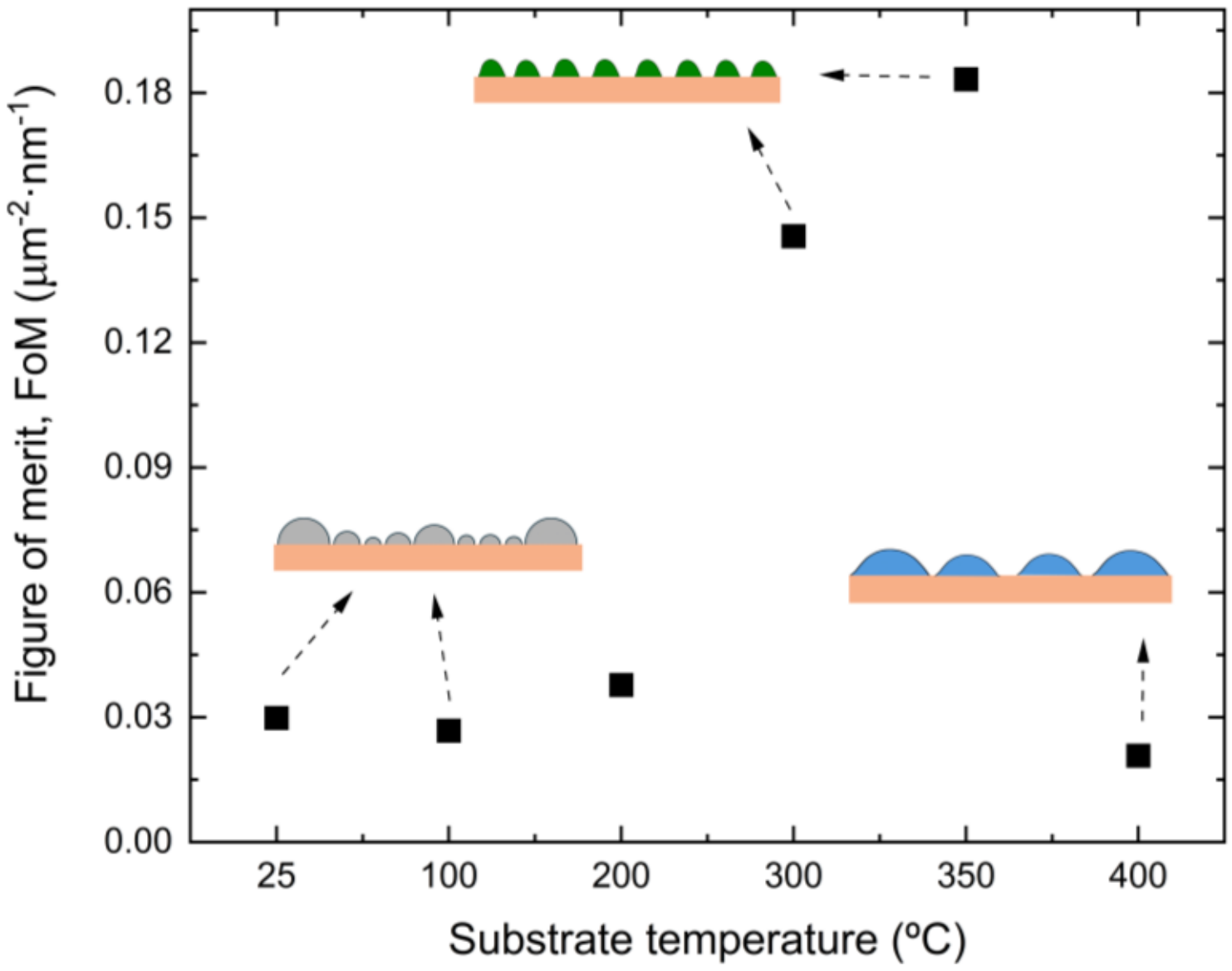

*Figure 4. Figure of merit (FoM) quantifying NP density and size uniformity as a function of substrate temperature according to equation 1. Sketches of the three different identified scenarios are included.*

In order to demonstrate the correlation between size distribution and optical properties, reflectance spectra of Ga-NPs deposited during 80s were acquired for the three different scenarios of Fig. 4 (RT, 350 °C and 400 °C). The spectra of Figure 5 reveal the presence of two distinct localized surface plasmon resonance (LSPR) modes: a high-energy transversal mode in the ultraviolet, observed as a pronounced minimum in the reflectance curves, and a lower-energy longitudinal mode, visualized as a maximum spanning the visible to infrared (IR) region. The transversal mode corresponds to out-of-plane electron oscillations (perpendicular to the substrate), while the longitudinal mode arises from in-plane oscillations (parallel to the substrate). A sketch of each LSPRs mode in a simplified one-particle model has been included in Fig. 5 for clarification. The spectral position and sharpness of these LSPRs are severely sensitive to both the size and shape of the NPs, as well as their size distribution (homogeneity) within the array[11].

The energy position of each LSPR mode is closely correlated with the morphological evolution of the Ga-NPs as extracted from AFM analysis (see Fig. 1 and 2). Specifically, the minima corresponding to the transversal LSPR mode appears at 4.70, 4.84, and 4.59 eV for RT, 350 °C, and 400 °C growth, respectively, reflecting the trend in average NP heights in the dominant populations: the higher the NP, the lower the energy (longer wavelength) of the transversal mode, and vice versa. This alignment demonstrates the direct influence of vertical NP dimension on the out-of-plane LSPR. Simultaneously, the longitudinal LSPR maxima are positioned at 2.19 and 2.35 eV for RT and 350 °C, respectively, with the maximum shifting further to below 1.2 eV in the IR for the 400 °C sample out of our spectrophotometer range. The progressive redshift of this mode with increasing NP diameter again matches the diameter evolution (longitudinal dimension) observed in the histograms of Fig. 1. Notably, the energy difference between longitudinal and transversal modes is smallest for the 350 °C sample, consistent with its higher aspect ratio (height/diameter) determined in Fig. 2. The more hemispherical (less anisotropic and closer aspect ratio to 1) the NPs, the closer these two LSPRs mode approach, highlighting the interplay between NP shape and plasmonic behavior.

To quantify the optical performance of the Ga-NPs, the well-known[26] plasmonic quality factor (Q) was calculated for the in-plane LSPR mode as Q = LSPR energy/FWHM, with both parameters extracted from Gaussian fits to the experimental reflectance spectra. In the literature, it has been already demonstrated that this parameter can be enhanced by modifying the arrangement of the nanostructures with temperature in Au NPs[27]. However, in Ga-NPs it is especially important since LSPRs can be broad due to its intrinsic material damping and higher interband transitions in the VIS and IR region[28]. For our experimental results, at RT, a Q of 0.3 ± 0.1 is obtained, whereas for the NPs array grown at 350 °C, this value rises sharply to 0.84 ± 0.07. This enhancement is a direct result of the narrower LSPR achieved through improved NP size uniformity as expected from the high FoM previously established in Fig. 4. In contrast, the Q-factor at 400 °C could not be reliably evaluated, as the LSPR maximum shifts outside the measured spectral range. Interestingly, the optimized Q value for 350 °C is already closer to single-particle Q-factors of Ga-NPs that ranges from 1 to 3 from UV to IR region[14].

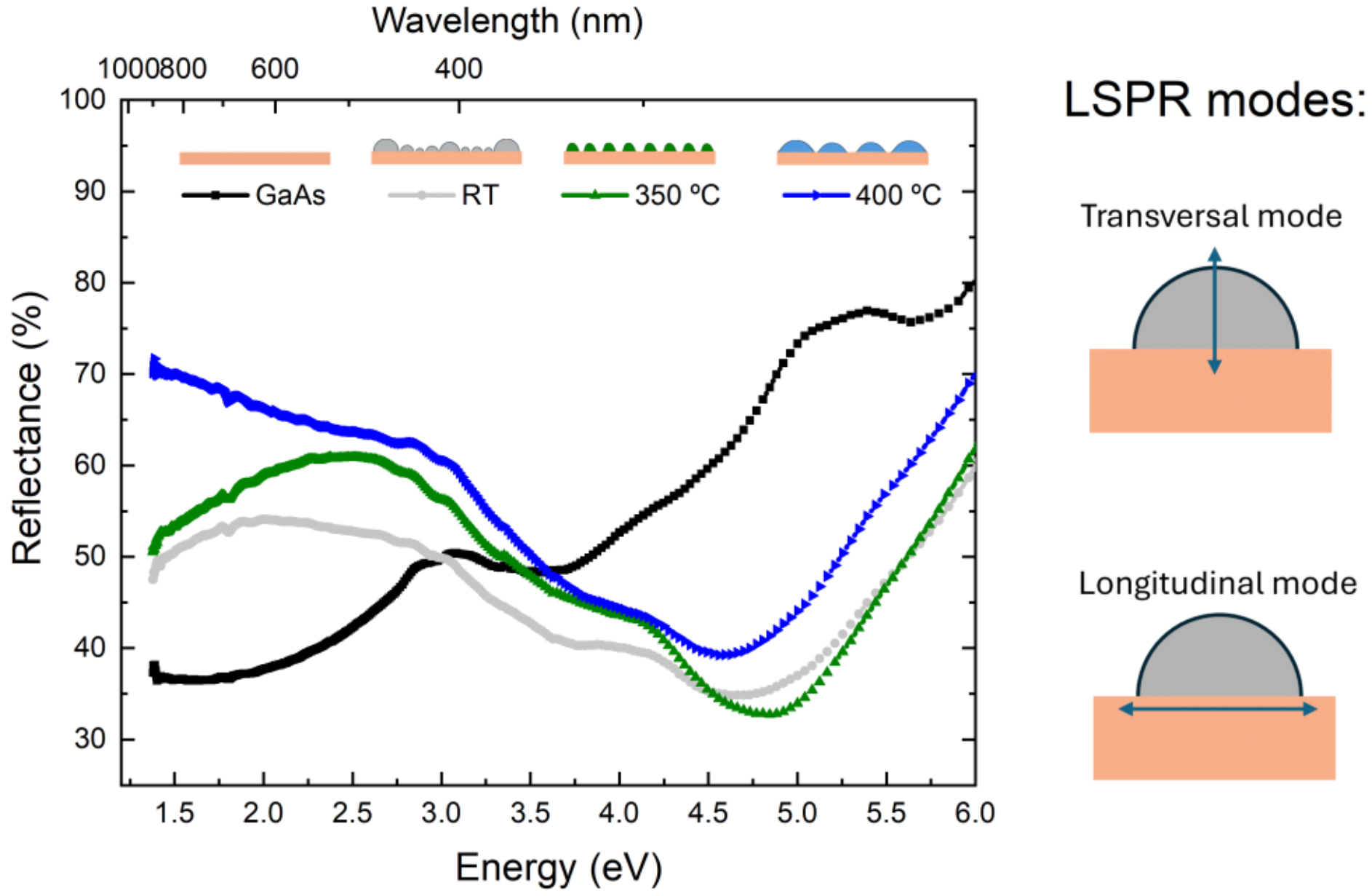

*Figure 5. Reflectance spectra of Ga-NP arrays grown at different substrate temperatures: RT, 350°C, and 400°C. Two main LSPR modes are identified: a transversal (out-of-plane) and a longitudinal (in-plane) mode observed as minimum and maximum in the graph. A schematic illustration is also included of those relative to the NP shape.*

Figure 6 addresses the crucial question of whether the temperature-driven size homogenization processes observed in previous experiments at small Ga-NP sizes remain robust and reproducible when extended to larger dimensions, that is a key requirement for practical scalability in applications. To this end, the same thermal evaporation protocol was carried out with a substantially increased Ga deposition time of 150s instead of 80s, producing NPs with larger final sizes. The morphological characterization by SEM at substrate temperatures of RT, 350 °C and 400 °C allows direct comparison with prior results in Fig. 1 at shorter deposition times.

As expected, a bimodal size distribution is obtained at RT deposition (Fig. 6 a) and d)). At 350 °C in Fig. 6 b), the SEM image reveals that the Ga-NP array undergoes the same pronounced transition toward uniformity as previously observed: the size distribution becomes narrower (size homogenization), and the NP density increases for a given mean diameter. This effect is again attributed to efficient surface diffusion and Ostwald ripening during growth, leading to the elimination of smaller particles in favor of a more monodisperse population. Notably, the average NP diameter actually decreases significantly in this regime compared to RT from 187 nm to 117 nm as observed in the histograms, confirming that increased substrate temperature promotes desorption of Ga and restricts the total mass available for NP formation. At the highest substrate temperature of 400 °C, diffusion effects become even more pronounced. The average NP size increases substantially, and the density drops dramatically as highly mobile adatoms favor the growth of fewer, larger NPs, while significant desorption further limits the number of NPs. Additionally, the SEM images indicate a detectable loss of NP sphericity, likely due to pronounced flattening or morphological relaxation at these elevated temperatures as anticipated for smaller NPs in Fig. 1 f) and Fig. S2.

To demonstrate the scalability of the temperature-driven homogenization mechanism to a wide range of NP sizes, we extended the study from 10s to 150s of deposition times at both RT and 350 °C, using SEM characterization. Their corresponding images and histograms are included in Figure S4. For every deposition time, samples grown at substrate temperature of 350 °C systematically show (i) higher NP density for a given mean diameter and (ii) narrower size distributions than RT samples, confirming the robustness of the mechanism from very small NPs (≈10-20 nm) up to large ones (≈190 nm). The figure of merit FoM (Eq. 1) calculated from these SEM data and shown in Figure S5 further quantifies this size scalability: 350 °C consistently yields higher FoM values across the entire deposition time range.

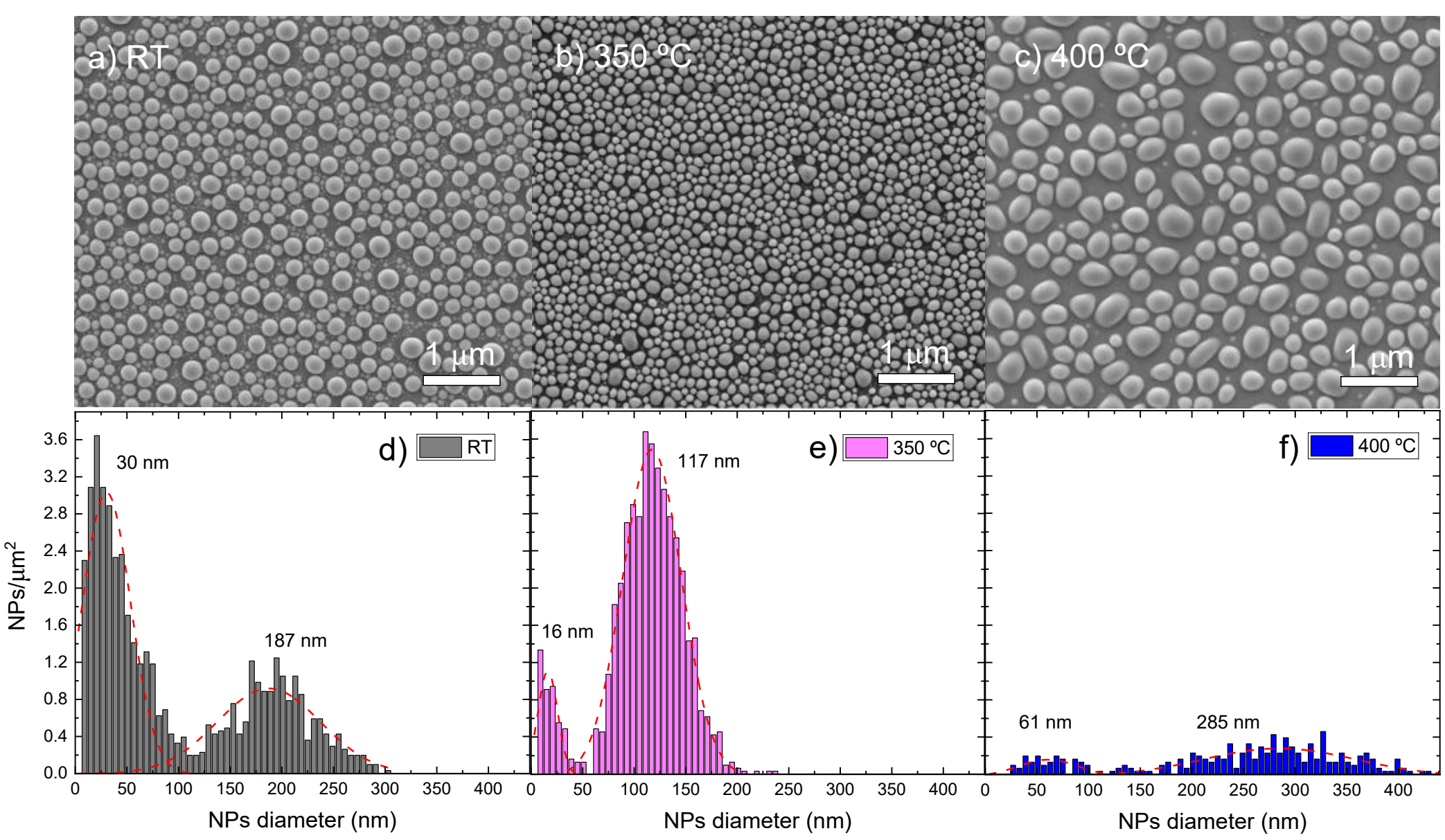

*Figure 6. Morphological comparison of larger Ga-NP arrays grown with extended deposition time. a) to c) SEM images depict Ga-NP arrays deposited for 150 seconds at substrate temperatures of RT, 350°C and 400°C, respectively. d) to f) corresponding histograms analyzed with ImageJ.*

Overall, Fig. 6, S4 and S5 robustly demonstrate that the key trends identified in Figure 1 and 2 at a Ga deposition time of 80s are fully scalable to larger and smaller NP systems. At moderate substrate temperature (350 ºC), a size homogenization, density increase and higher aspect ratio happened while increase in NP diameter and the onset of shape relaxation is observed at the highest temperature (400 ºC). This validates the size scalability of the temperature-driven size homogenization process, affirming its applicability for the fabrication of dense and uniform Ga-NP arrays across a broad range of particle sizes for advanced functional applications.

The temperature-driven method described here for achieving uniform Ga-NP arrays is prospective applicable to a variety of substrates, though the optimal temperature range will depend on each substrate's chemical composition, crystallography, and surface energy. Numerous studies have investigated Ga diffusion and growth dynamics on different materials, including sapphire, Si, GaAs, $SiN_x$ and patterned oxide surfaces, as well as polymer or flexible substrates[9, 12, 16, 29-31]. These works consistently show that the surface diffusion coefficient of Ga adatoms-crucial for NP nucleation and ordering-varies with substrate chemistry and morphology, surely impacting the onset of uniform NP growth and the temperature required to achieve enhanced dispersity. While it may be necessary to adjust the substrate temperature regime for different surfaces, we believe that the underlying principle of thermally activated surface diffusion as a driver for NP size homogenization remains valid to other systems. Thus, with careful calibration of growth parameters, this strategy can be extended to engineer dense, uniform Ga-NP assemblies on a wide range of technological platforms.

## Conclusions

We have demonstrated that substrate temperature critically governs the size distribution, ordering, and morphological evolution of Ga-NP arrays grown by simple Joule-effect thermal evaporation. An optimal temperature range around 300-350 °C facilitates the transformation from polydisperse, bimodal ensembles to dense spatially uniform NP arrays, corresponding to enhanced plasmonic performance. Above this window, increased atomic mobility and desorption lead to NP flattening, reduced density, and broader size distributions. In-situ post-growth annealing establishes Ostwald ripening as the primary coarsening mechanism at elevated temperatures and emphasizes the importance of rapid cooling and oxide shell formation to maintain the optimized NP distribution. TEM with EELS analysis confirms the preservation of the core-shell architecture at elevated substrate growth temperatures and validates the liquid character of the Ga core. Optical characterization reveals distinct plasmonic modes whose energies and quality factors correlate strongly with NP dimensions and improved size uniformity. Size scalability tests confirm that these temperature-driven enhanced spatial uniformity principles extend to a wide range of NP sizes, underscoring the robustness of the synthesis approach. Collectively, our findings offer practical pathways to control NP uniformity and plasmonic properties through thermal processing, guiding the design of future plasmonic and photonic devices based on Ga nanostructures.

## Data statement

The authors declare that all data supporting the findings of this study are available at https://doi.org/10.5281/zenodo.17989223.

## Acknowledgments

This research was supported by Proyect FEDER-UCA-2024- A2-08 funded by Programa Operativo FEDER Andalucía 2021-2027 and by Consejería de Universidad, Investigación e Innovación, Junta de Andalucía. It has been also supported by the Madrid Government (Comunidad de Madrid-Spain) under the Multiannual Agreement 2023-2026 with Universidad Politécnica de Madrid in the Line A, Emerging PhD researchers through the project CADENCE. Authors acknowledge the use of instrumentation as well as the technical advice provided by the National Facility ELECMI ICTS, node of Division de Microscopía Electrónica (DME) at Universidad de Cádiz from ELECMI 2024 call (ref: ELC667-2025). ICTS Micronanofabs is also acknowledged.

# References

1. Song, H. *et al*. Ga-Based Liquid Metal Micro/Nanoparticles: Recent Advances and Applications. *Small* **16**, e1903391 (2020).

2. Tian, H. *et al*. Recent advances for core–shell gallium-based liquid metal particles: properties, fabrication, modification, and applications. *Nanoscale* **17**, 11934–11959 (2025).

3. Chiew, C., Morris, M. J. & Malakooti, M. H. Functional liquid metal nanoparticles: synthesis and applications. *Mater. Adv.* **2**, 7799–7819 (2021).

4. Liu, P. Q., Miao, X. & Datta, S. Recent advances in liquid metal photonics: technologies and applications [Invited]. *Opt. Mater. Express* **13**, 699–727 (2023).

5. Sun, X. & Li, H. Recent progress of Ga-based liquid metals in catalysis. *RSC Advances* **12**, 24946–24957 (2022).

6. Du, S. *et al*. Emerging frontiers in biomedicine: a bibliometric analysis of gallium-based nanomaterials. *Coordination Chemistry Reviews* **544**, 216966 (2025).

7. Catalán-Gómez, S. *et al*. Gallium nanoparticles as antireflection structures on III-V solar cells. *Solar Energy Materials and Solar Cells* **265**, 112632 (2024).

8. Catalán-Gómez, S., Redondo-Cubero, A., Palomares, F. J., Nucciarelli, F. & Pau, J. L. Tunable plasmonic resonance of gallium nanoparticles by thermal oxidation at low temperatures. *Nanotechnology* **28**, 405705 (2017).

9. Losurdo, M., Suvorova, A., Rubanov, S., Hingerl, K. & Brown, A. S. Thermally stable coexistence of liquid and solid phases in gallium nanoparticles. *Nature Mater* **15**, 995–1002 (2016).

10. Dickey, M. D. Stretchable and soft electronics using liquid metals. *Adv. Mater* **29** (2017).

11. Catalán-Gómez, S. *et al*. Plasmonic coupling in closed-packed ordered gallium nanoparticles. *Sci Rep* **10**, 4187 (2020).

12. Sahu, R. R. *et al*. Single-step fabrication of liquid gallium nanoparticles via capillary interaction for dynamic structural colours. *Nat. Nanotechnol.* **19**, 766–774 (2024).

13. Catalán-Gómez, S. *et al*. Modification of the Mechanical Properties of Core-Shell Liquid Gallium Nanoparticles by Thermal Oxidation at Low Temperature. *Part. Part. Syst. Charact.* **38**, 2100141 (2021).

14. Horák, M., Čalkovský, V., Mach, J., Křápek, V. & Šikola, T. Plasmonic Properties of Individual Gallium Nanoparticles. *J. Phys. Chem. Lett.* **14**, 2012–2019 (2023).

15. Lee, S. A. & Link, S. Chemical Interface Damping of Surface Plasmon Resonances. *Acc. Chem. Res.* **54**, 1950–1960 (2021).

16. Baraissov, Z., Panciera, F., Travers, L., Harmand, J. & Mirsaidov, U. Growth Dynamics of Gallium Nanodroplets Driven by Thermally Activated Surface Diffusion. *J. Phys. Chem. Lett.* **10**, 5082–5089 (2019).

17. Sherman, Z. M., Milliron, D. J. & Truskett, T. M. Distribution of Single-Particle Resonances Determines the Plasmonic Response of Disordered Nanoparticle Ensembles. *ACS Nano* **18**, 21347–21363 (2024).

18. Schenk, F. *et al*. Rational Design for
Monodisperse Gallium Nanoparticles
by In Situ Monitoring with Small-Angle X-ray Scattering. *J Am Chem Soc* **147**, 12105–12114 (2025).

19. Xu, X. *et al*. Site-controlled fabrication of Ga nanodroplets by focused ion beam. *Appl. Phys. Lett.* **104**, 133104 (2014).

20. Canniff, J. C., Jeon, S., Huang, S. & Goldman, R. S. Formation and coarsening of near-surface Ga nanoparticles on SiNx. *Appl. Phys. Lett.* **106**, 243102 (2015).

21. Yan, J. *et al*. Shape-controlled synthesis of liquid metal nanodroplets for photothermal therapy. *Nano Res.* **12**, 1313–1320 (2019).

22. Bogar, M. *et al*. Interplay Among Dealloying, Ostwald Ripening, and Coalescence in PtXNi100–X Bimetallic Alloys under Fuel-Cell-Related Conditions. *ACS Catal.* **11**, 11360–11370 (2021).

23. Košutová, T. *et al*. Temperature-Driven Morphological and Microstructural Changes of Gold Nanoparticles Prepared by Aggregation from the Gas Phase. *ACS Omega* **10**, 22052–22061 (2025).

24. Goodman, E. D. *et al*. Size-controlled nanocrystals reveal spatial dependence and severity of nanoparticle coalescence and Ostwald ripening in sintering phenomena. *Nanoscale* **13**, 930–938 (2021).

25. Tian, Y. *et al*. Fast coalescence of metallic glass nanoparticles. *Nat Commun* **10**, 5249 (2019).

26. Bosman, M. *et al*. Surface Plasmon Damping Quantified with an Electron Nanoprobe. *Sci Rep* **3**, 1312 (2013).

27. Handoyo, T., Firmansyah, T. & Kondoh, J. The quality factor enhancement on gold nanoparticles film for localized surface plasmonic resonance chip sensor. *e-Prime - Advances in Electrical Engineering, Electronics and Energy* **7**, 100406 (2024).

28. Gutiérrez, Y., Brown, A. S., Moreno, F. & Losurdo, M. Plasmonics beyond noble metals: Exploiting phase and compositional changes for manipulating plasmonic performance. *J. Appl. Phys.* **128**, 080901 (2020).

29. Sobanska, M. *et al*. Surface Diffusion of Gallium as the Origin of Inhomogeneity in Selective Area Growth of GaN Nanowires on AlxOy Nucleation Stripes. *Crystal Growth & Design* **20**, 4770–4778 (2020).

30. Peng Xu, D. *et al*. Gallium diffusion through cubic GaN films grown on GaAs(1 0 0) at high-temperature using low-pressure MOVPE. *Journal of Crystal Growth* **191**, 646–650 (1998).

31. Apostolopoulos, V., Hickey, L. M. B., Sager, D. A. & Wilkinson, J. S. Diffusion of gallium in sapphire. *Journal of the European Ceramic Society* **26**, 2695–2698 (2006).

# Supporting Information of:

# Thermal Control of Size Distribution and Optical Properties in Gallium Nanoparticles

S. Catalán-Gomez[1,2]*, M. Ibáñez[1], J. Rico[1], V. Braza[3], D. F. Reyes[3], M. Villanueva-Blanco[1], E. Squiccimarro[1,#] and J.M. Ulloa[1]

[1]Instituto de Sistemas Optoelectrónicos y Microtecnología, Universidad Politécnica de Madrid, ETSI Telecomunicación, Av. Complutense 30, E-28040 Madrid, Spain

[2]Departamento de Ingeniería Eléctrica, Electrónica Automática y Física Aplicada, Universidad Politécnica de Madrid, ETSI y Diseño Industrial, Ronda de Valencia 3, E-28012 Madrid, Spain

[3]Departamento de Ciencia de los Materiales e IM y QI, Universidad de Cádiz, E-11510 Puerto Real, Cádiz, Spain

*Corresponding author: sergio.catalan.gomez@upm.es

#Current address: Department of Applied Science and Technology, Polytechnic of Turin, Corso Duca degli Abruzzi 24, 10129 Turin, Italy

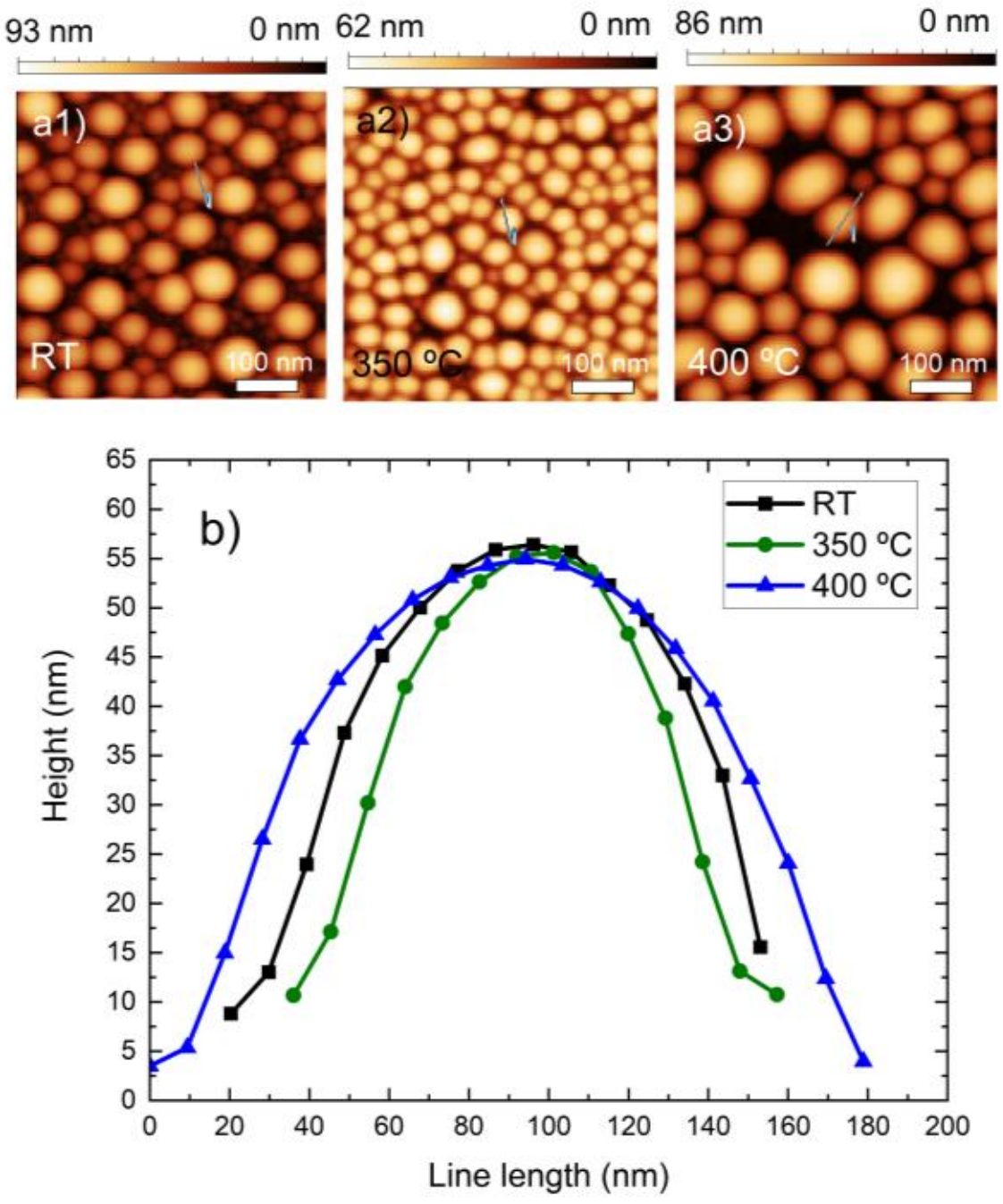

*Figure S1. a1)-a3) 1x1 µm² AFM images of Ga-NPs deposited on GaAs at substrate temperature of RT, 350 °C, and 400 °C, respectively. b) AFM height profiles of selected Ga-NPs) from AFM images a1-a3).*

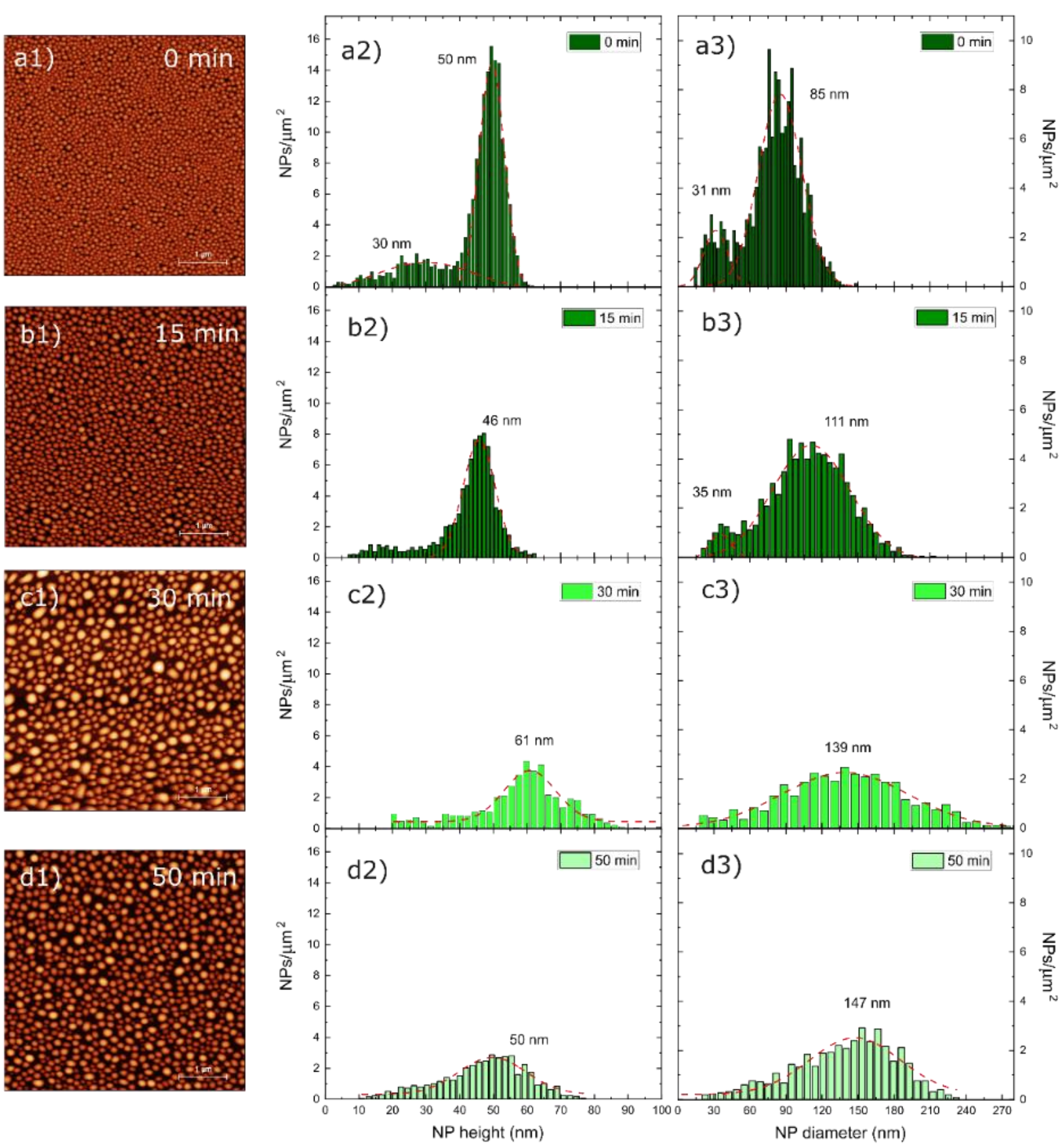

*Figure S2. 5x5 µm2 AFM images of Ga-NPs growth at substrate temperature of 350 °C but maintaining the sample for different times within the vacuum chamber at 350 °C in the absence of additional Ga flux. All AFM images have the same vertical axis from 0 to 100 nm. Second and third columns correspond to height and diameter distribution histograms from AFM images of first column, respectively.*

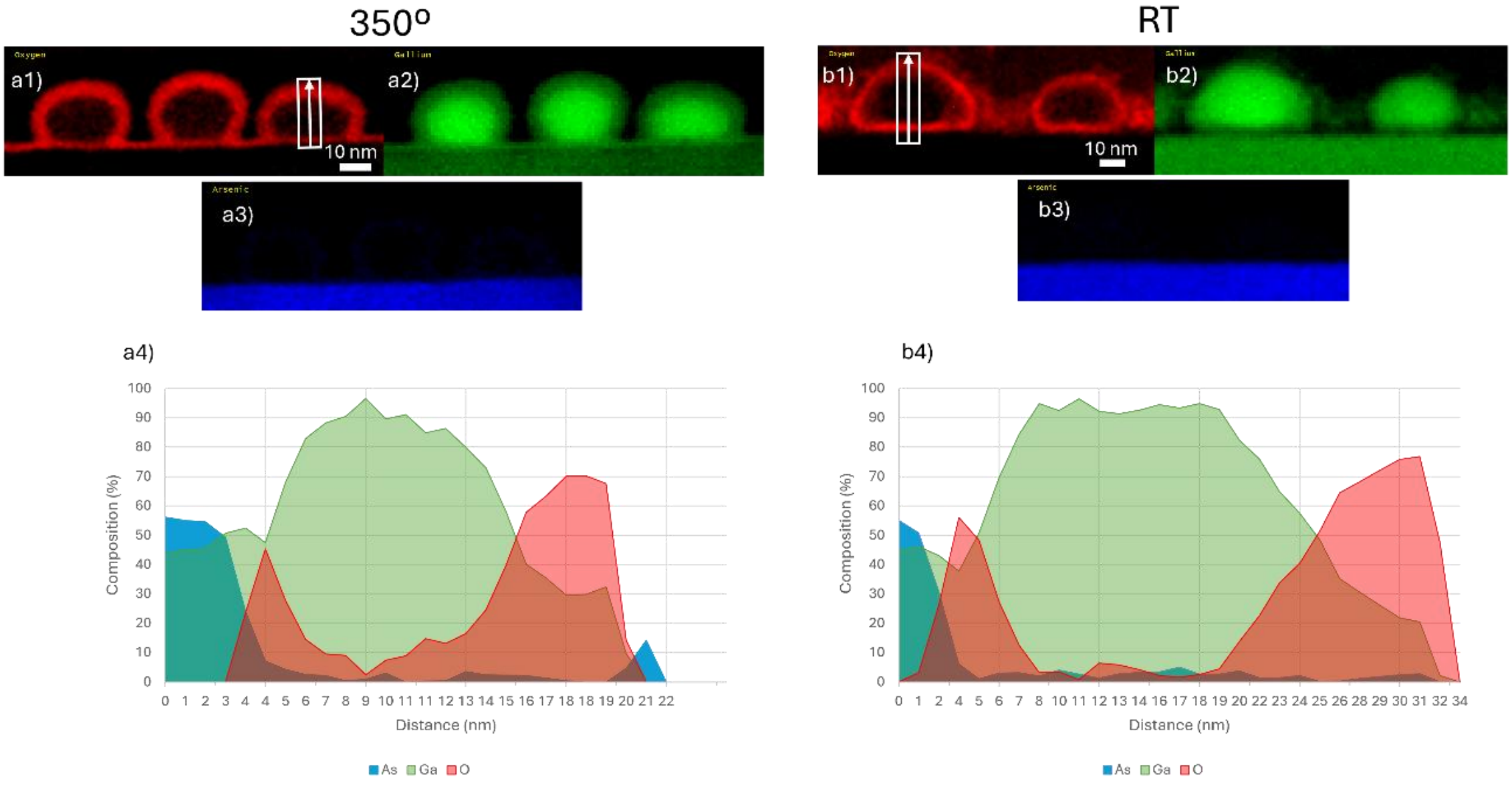

*Figure S3. EELS elemental maps and compositional profiles of Ga-NPs. Column a) and b) correspond to the RT sample, and the 350 °C sample, respectively. In both columns, elemental maps of O, Ga, and As, are shown. a4) and b4) show the elemental profiles through the NPs along the growth direction indicated in a1) and b1).*

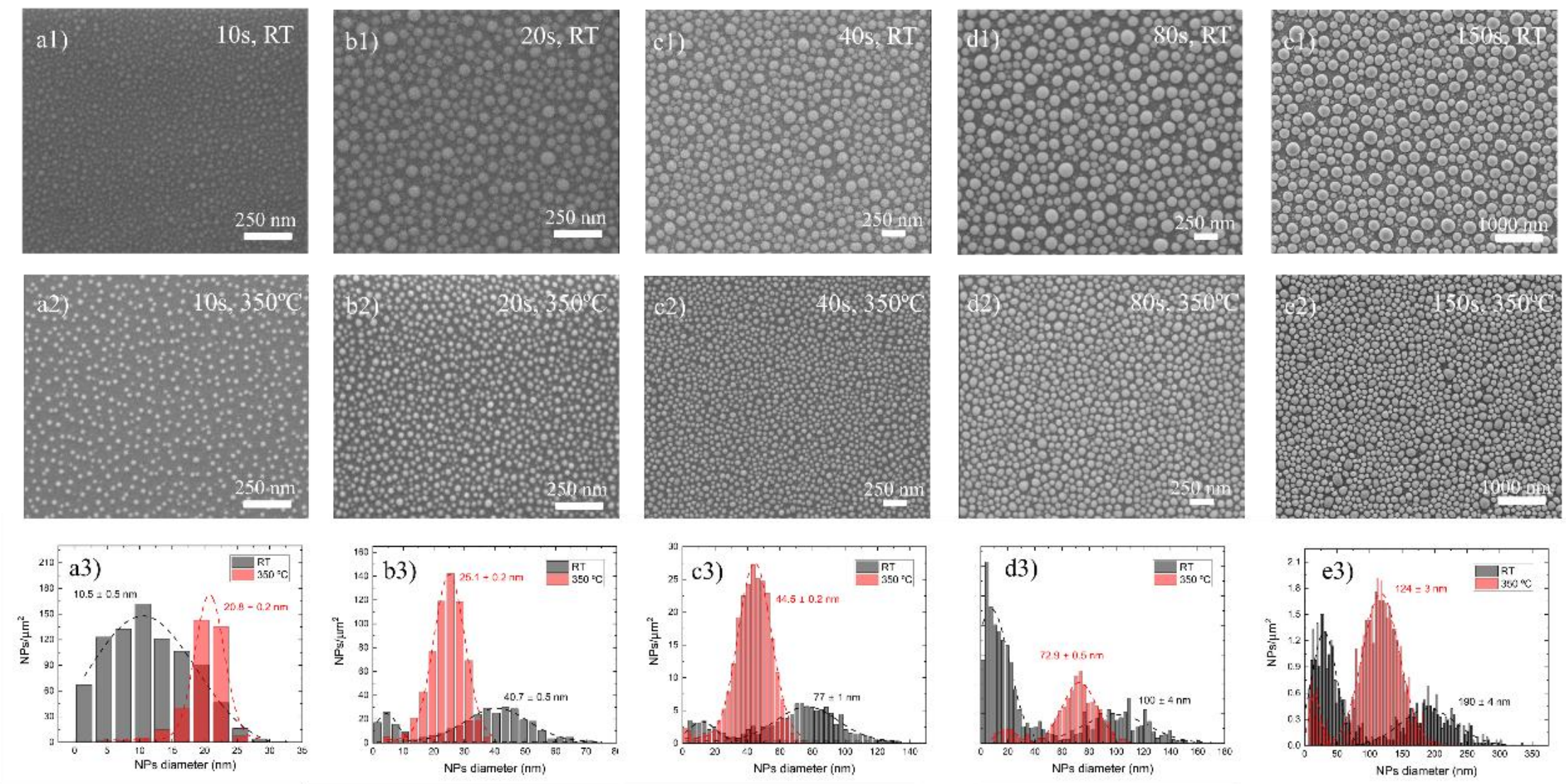

*Figure S4. Scalability of size homogenization mechanism to a broad deposition time. (a–e) Representative SEM images and Gaussian-fitted diameter histograms for Ga deposition times of 10, 20, 40, 80 and 150s at RT (first row) and 350 °C (second row) substrate temperatures. For all times, 350 °C samples exhibit higher density and narrower distributions than RT, demonstrating size scalability of the homogenization effect.*

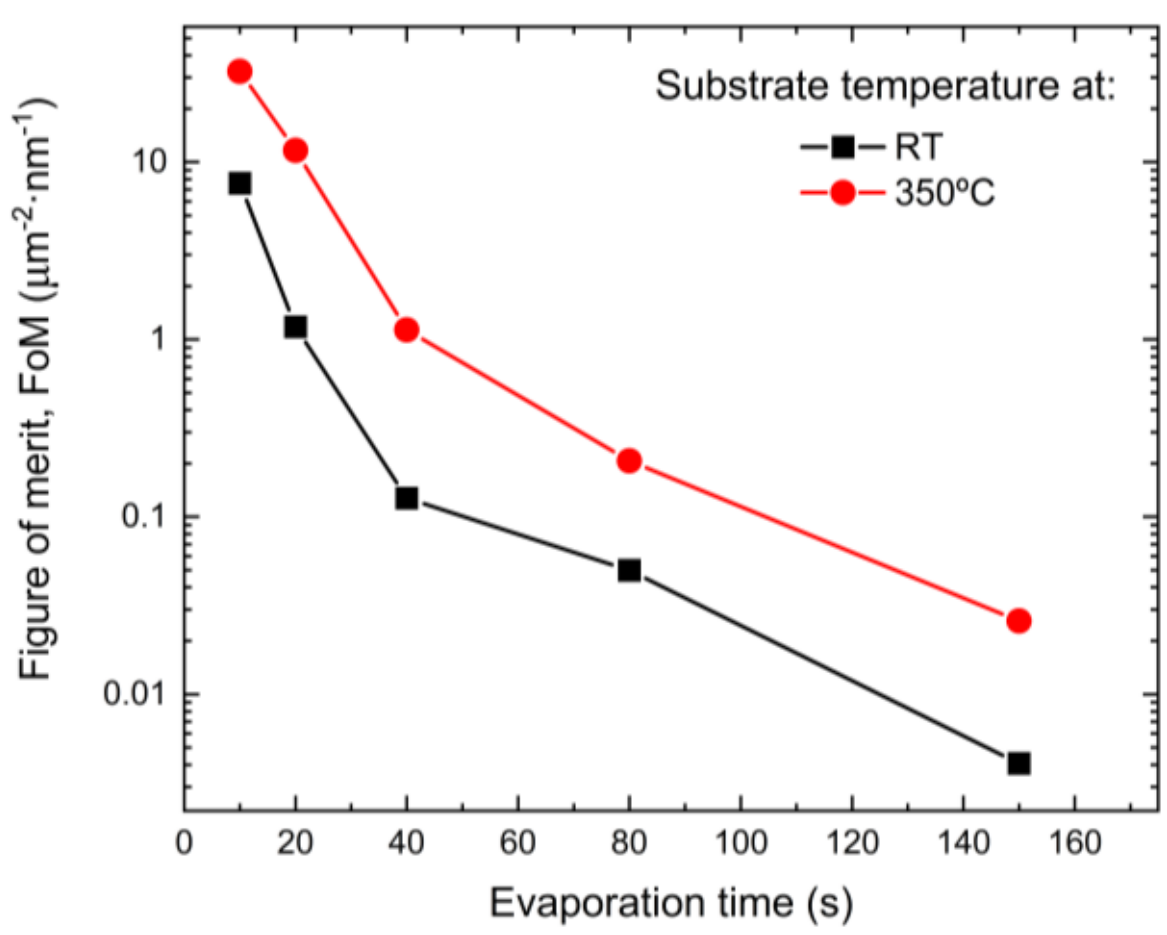

*Figure S5. Figure of merit (FoM = NP density / size FWHM) as a function of evaporation time for RT (black squares) and 350 °C (red circles) substrate temperatures. FoM values are calculated from Gaussian fits to SEM diameter histograms (Fig. S4) using the same procedure as in Fig. 4. Across the full range (10-150 s), 350 °C consistently yields higher FoM, confirming systematic scalability with NP size.*